\documentclass[12pt,epsfig]{article}

\usepackage{amsfonts,amssymb}

\newcommand{\beq}{\begin{equation}}
\newcommand{\eeq}{\end{equation}}
\newcommand{\beqa}{\begin{eqnarray}}
\newcommand{\eeqa}{\end{eqnarray}}
\newcommand{\reef}[1]{(\ref{#1})}

\def\p{\partial}

\let\bar=\overbar

\def\p4n{{\mbox{\scriptsize (4+N)}}}
\def\p4{{\mbox{\scriptsize (4)}}}

\begin{document}

\setlength{\unitlength}{1mm}

\thispagestyle{empty} \rightline{\small DCPT-03/31\hfill}
\rightline{\small hep-th/0306267\hfill}
\vspace*{2cm}

\begin{center}
{\bf \LARGE Topological charges for branes }\\[.5em]
{\bf \LARGE in M-theory}
\vspace*{1cm}

{\bf Emily~J.~Hackett-Jones}\footnote{E-mail: {\tt
e.j.hackett-jones@durham.ac.uk}}, {\bf
David~C.~Page}\footnote{E-mail: {\tt page@physics.utoronto.ca}}
{\bf and Douglas~J.~Smith}\footnote{E-mail: {\tt
douglas.smith@durham.ac.uk}}

\vspace*{0.3cm} \vspace{0.5cm}
$^{1,3}$Centre for Particle Theory \\
Department of Mathematical Sciences \\
University of Durham, Durham DH1 3LE, U.K. \\
\vspace{0.5cm}
$^{2}$Department of Physics and Astronomy,\\
University of Toronto, Canada. \\

\vspace{2cm}
{\bf ABSTRACT}
\end{center}

We propose a simple form for the superalgebra of M2 and M5-brane
probes in arbitrary supersymmetric backgrounds of 11d
supergravity, extending previous results in the literature. In
particular, we identify the topological
 charges in the algebras and find BPS bounds for the energies. The charges are given by the integral over a brane's spatial
 worldvolume of a certain closed form built out of the Killing spinors and background
 fields. The existence of such closed forms for arbitrary supersymmetric backgrounds
 generalises  the existence of calibration forms for special holonomy manifolds.

\vfill
\setcounter{page}{0}
\setcounter{footnote}{0}
\newpage


\section{Introduction}

It has been known for some time that the supersymmetry algebras of brane worldvolume theories can contain topological charges which extend the spacetime supersymmetry algebra \cite{deAzcarraga}. As a simple example consider a supermembrane probe in flat eleven dimensional spacetime.

In this case the spacetime superalgebra is the 11d
super-Poincar\'{e} algebra. If we couple a supermembrane probe to
this background, the resulting membrane action inherits the
symmetries of the background, but with a modification to the
supertranslation algebra \cite{deAzcarraga,Townsend:m-theory}:

\beq \label{brane1} \{ Q_\alpha, Q_\beta \} = (C
\Gamma^M)_{\alpha\beta} P_{M} \pm \frac{1}{2} (C \Gamma_{M
N})_{\alpha\beta} Z^{M N}, \eeq where \beq Z^{M N} = \int dX^M
\wedge dX^N \eeq and the integration is taken over the spatial
worldvolume of the membrane\footnote{Here and in the following,
when we write expressions involving forms defined on the
background being integrated over the brane worldvolume, a pullback
is implied. }.

$Z^{M N}$ is the integral of a closed form and so the second term
in the membrane supertranslation algebra \reef{brane1} depends
only on the homology class of the configuration. The existence of
such a topological charge in the superalgebra allows massive
objects such as branes, which carry the charge, to have
supersymmetric ground states.

Subsequent studies have extended this analysis to branes of
various types in other fixed supersymmetric backgrounds, see e.g.
\cite{Sorokin, Sato}. In \cite{Gutowski} an expression for the membrane and
fivebrane superalgebras was given which is valid for a more
general class of backgrounds. Specifically, the analysis of
\cite{Gutowski} is relevant to backgrounds which have a timelike
Killing vector appearing in the supertranslation algebra and which
have certain implicit restrictions on the background fields. A
further paper \cite{Barwald} studies the M5-brane superalgebra in
the presence of non-zero worldvolume fields. The present work
proposes a simple and natural generalisation of these results to
arbitrary supersymmetric backgrounds and worldvolume fields.

The outline of the paper is as follows. In section
\ref{closedforms} we review some recent work on the classification
of supersymmetric solutions of 11d supergravity
\cite{Gauntlett:geometry}. An important step in this
classification is the construction of certain forms built out of
the Killing spinors which obey differential relations descended
from the Killing spinor equations. By simple manipulations we show
that these relations are equivalent to the existence of certain
closed forms which we will later argue to be the topological
charges for branes.

In section \ref{spacetime} we review the construction of the
spacetime superalgebra associated with a supergravity background
\cite{Figueroa}. This will help us to fix some useful notations
and will suggest a natural construction for the brane
superalgebras.

In section \ref{membranes} we consider the superalgebra for a
membrane probe and in section \ref{fivebranes}, that of an
M5-brane. We make a natural proposal for the form of the
superalgebras which agrees with the known examples and makes use
of the closed forms constructed in section \ref{closedforms}. In
section \ref{examples}, we give an example of the construction
applied to the supergravity background sourced by an M5-brane. We
conclude with a short discussion.

We note that whilst this paper was being completed
\cite{Martelli:g-structures} appeared which has some overlap with
the current work. In particular our expression \reef{BPS} for the
BPS bound on the energy/momentum of the fivebrane agrees with
equation (4.6) of \cite{Martelli:g-structures}.


\section{Killing spinors and closed forms}
\label{closedforms}

Recently a great deal of progress has been made in understanding
the general structure of supersymmetric solutions of supergravity
theories
\cite{Gauntlett:geometry,Gauntlett:all,Gauntlett:all2,Gauntlett:superstrings,Duff:hidden,Hull,Papadopoulos}.
We shall be interested in the case of 11 dimensional supergravity
which was investigated by Gauntlett and Pakis
\cite{Gauntlett:geometry}. They studied the consequences of having
a spinor field which obeys the Killing spinor equations:

\beq \label{Killing} \tilde{D}_M \epsilon = 0 \eeq where \beq
\label{Killing2}\tilde{D}_M \epsilon \equiv \nabla_M \epsilon +
\frac{1}{288} \left[ \Gamma_M^{\, \, \, \,N P Q R} - 8
\delta_M^{N} \Gamma^{P Q R} \right]F_{N P Q R} \epsilon
 \eeq
and $F$ is the four-form field strength of 11d supergravity.

In order to study the consequences of \reef{Killing}, it is helpful to repackage  $\epsilon(x)$ in terms of the following one, two and five-forms:
\beqa
K_M &=& \bar{\epsilon} \Gamma_M \epsilon \nonumber \\
\omega_{M N} &=& \bar{\epsilon} \Gamma_{M N} \epsilon \nonumber \\
\Sigma_{M N P Q R} &=& \bar{\epsilon} \Gamma_{M N P Q R} \epsilon \, .
\eeqa

It is straightforward to check that the zero, three and four-forms
built in a similar way vanish because of the antisymmetry of the
relevant $\Gamma$ matrices. Also $\epsilon(x)$ can be
reconstructed (up to a sign) from knowledge of $K, \omega$ and
$\Sigma$. This follows from the completeness of the $\Gamma$
matrices.

It is not true, however, that an arbitrary set of one, two and five forms are equivalent to a spinor - rather there are algebraic relations between them which follow from Fierz identities. A better way to find these relations uses some algebraic facts about spinors in 10+1 dimensions discussed in \cite{Bryant}.

Spinors form a representation of $Spin(1,10)$ and a natural question to ask is what are the possible orbits of a spinor under $Spin(1,10)$. Clearly $K^2 = K^\mu K_\mu$ is a Lorentz scalar and thus is the same along orbits of the group. In fact there are no other independent invariants and $Spin(1,10)$ acts transitively on the level sets of $K^2$ \cite{Bryant} \footnote{Except that the zero spinor is an obvious fixed point. }.

Furthermore, the only possibilities for $K^2$ are $K^2 < 0$ or
$K^2 = 0$, i.e.\ $K^\mu$ is either  timelike or null. We can choose a
convenient normal form for a spinor of either type and deduce that
any other such spinor is related by a Lorentz transformation (and
possible rescaling if $K^2 < 0$.) Working in this way gives a
rather efficient way to define (Lorentz and scaling covariant)
identities between the forms $K, \omega$ and $\Sigma$.

Consider first the case in which $K^2 < 0$. A possible set of
projection conditions which define the spinor $\epsilon$ is given
as follows: \[ \Gamma_{0 1 2} \epsilon = \Gamma_{0 3 4} \epsilon =
\Gamma_{0 5 6} \epsilon = \Gamma_{0 7 8} \epsilon = \Gamma_{0 9
\natural} \epsilon =  \epsilon \] \beq \label{timelike} \Gamma_{0
1 3 5 7 9} \epsilon = \epsilon \, .\eeq These provide a set of
five independent\footnote{Note that one of the six projections
given is satisfied automatically as a consequence of the other
five and the fact that $\Gamma_{0123456789\natural} \equiv 1$.},
commuting projections
 and thus determine a
unique spinor up to scale. The scale of the spinor is given by
fixing \beq \epsilon^T \epsilon = \Delta \eeq Using the projection
conditions \reef{timelike} we can uniquely determine the forms $K,
\omega$ and $\Sigma$: \[ K = \Delta e^0 \]
\[\omega = \Delta (e^1 \wedge e^2 + e^3 \wedge e^4 + e^5 \wedge
e^6 + e^7 \wedge e^8 + e^9 \wedge e^{\natural}) \] \beq
\label{su5structure}\Sigma = \frac{1}{2} \Delta^{-2} K \wedge
\omega \wedge \omega + \Delta Re(\Omega) \, ,\eeq where $\Omega$
is the holomorphic 5-form: \beq \Omega = (e^1 + ie^2) \wedge (e^3
+ i e^4) \wedge (e^5 + i e^6) \wedge(e^7 + i e^8) \wedge(e^9 + i
e^{\natural}) \, .\eeq Note that $K$ is indeed a timelike vector
for this choice of projections and the forms define an $SU(5)$
structure corresponding to the stability group of $\epsilon$. Now,
since $Spin(1,10)$ acts transitively on the level sets of $K^2$,
we can bring the projection conditions for {\it any} spinor with
$K^2 < 0$ into the standard form given above, by an appropriate
choice of vielbein. Thus, a spinor with $K^2 < 0$ is equivalent to
a set of forms of the type listed in equation \reef{su5structure}.

Now consider the case in which $K$ is null. A possible set of
projection conditions for the spinor $\epsilon$ is given in this
case by: \[ \Gamma_{01} \epsilon = - \epsilon \] \beq \label{null}
\Gamma_{2345} \epsilon = \Gamma_{2367} \epsilon =\Gamma_{2389}
\epsilon =\Gamma_{2468} \epsilon = - \epsilon \, . \eeq Note that
using the identity $\Gamma_{0123456789\natural} \equiv 1$ we can
show that our projectors imply:\beq \Gamma_{\natural} \epsilon = -
\epsilon \, . \eeq We may use the fact that Lorentz
transformations act transitively on the spinors with fixed $K^2$
to always choose a vielbein such that the projections are of the
standard form given in equation \reef{null} whenever $K$ is null.

Having fixed the vielbein and the projection conditions we can
determine the forms $K, \omega$ and $\Sigma$ uniquely to be: \[ K
= \Delta (e^0 + e^1) \equiv \Delta e^+\] \[ \omega = - K \wedge
e^{\natural}\] \beq \Sigma = - K \wedge \phi \, ,\eeq where $\phi$
is the Cayley four-form: \beqa \phi = & & e^2 \wedge e^3 \wedge e^4\wedge e^5 +  e^6\wedge e^7 \wedge e^8 \wedge e^9 + e^2\wedge e^3 \wedge e^6 \wedge e^7- e^2\wedge e^5 \wedge e^6 \wedge e^9\nonumber\\
  &-& e^3\wedge e^4 \wedge e^7 \wedge e^8 + e^2\wedge e^4 \wedge e^6 \wedge e^8 + e^3\wedge e^5 \wedge e^7 \wedge e^9  + e^4\wedge e^5 \wedge e^8 \wedge e^9\nonumber \\
   &+& e^4\wedge e^5 \wedge e^6 \wedge e^7  - e^3\wedge e^4 \wedge e^6 \wedge e^9 + e^2\wedge e^3 \wedge e^8 \wedge e^9 - e^2\wedge e^5 \wedge e^7 \wedge
  e^8 \nonumber \\
  &-& e^2\wedge e^4 \wedge e^7 \wedge e^9 - e^3\wedge e^5 \wedge e^6
\wedge e^8 \, . \eeqa Note that $K$ is indeed null. We could set
$\Delta \equiv \epsilon^T \epsilon = 1$ by a boost in the $e^+
\equiv e^0 + e^1$ direction but it will  be convenient not to do
so. The forms $K, \omega$ and $\Sigma$ define a $(Spin(7) \ltimes
\mathbb{R}^8) \times \mathbb{R}$ structure \cite{Acharya:planes2}
corresponding to the stability group of the spinor $\epsilon$
\cite{Bryant}.

Now, we turn to the differential equations which the forms satisfy
as a consequence of $\epsilon(x)$ being a Killing spinor.
Expressions for the covariant derivatives of the forms $K, \omega$
and $\Sigma$ are given in equation (2.16) of
\cite{Gauntlett:geometry}. We shall be primarily interested in the
set of equations for the exterior derivatives of the forms: \beqa
dK &=& \frac{2}{3} \iota_\omega F + \frac{1}{3} \iota_\Sigma *F \label{dK}\\
d\omega &=& \iota_K F  \label{omega} \\
d \Sigma &=& \iota_K * F - \omega \wedge F \label{sigma} \eeqa and
the fact that $K^\mu$ is a Killing vector. These equations follow
as a result of the Killing spinor equations. Conversely, at least
in the case in which $K$ is timelike, this apparently weaker set
of equations, supplemented by the algebraic conditions on the
forms and the fact that $dF = 0$, is actually strong enough to
imply the full set of Killing spinor equations. This surprising
fact is highlighted in \cite{Gauntlett:geometry}. It would
certainly be interesting to find out if this fact generalises to
the case in which $K$ is a null Killing vector.

Now we show that these conditions imply the existence of a certain
closed 2-form and a closed 5-form on the manifold. We make
repeated use of the following identity. Let $X$ be a vector and
$\alpha$ a $p$-form. Then  \beq \mathcal{L}_X \alpha = d (\iota_X
\alpha) + \iota_X d\alpha \, , \eeq where $ \mathcal{L}_X$ denotes
the Lie derivative in the $X$ direction. Now applying $d$ to
equation \reef{omega} and using $dF = 0$ we find: \beq
{\mathcal{L}}_K F = 0 \, . \eeq and so $K$ generates a symmetry of
the solution. Let $A$ be a 3-form gauge potential for $F$, $dA =
F$. Then we can pick a gauge for $A$ which also preserves the
symmetry: \beq {\mathcal{L}}_K A = 0 \, . \eeq Now consider the
2-form $\omega + \iota_K A$. This is closed since: \beq d (\omega
+ \iota_K A) = \iota_K F + {\mathcal{L}}_K A - \iota_K F = 0. \eeq
We shall see that this closed 2-form is a natural choice for a
topological charge in the membrane superalgebra.

Now we construct a closed 5 form to play the same role in the fivebrane superalgebra. Since K is Killing and ${\mathcal{L}}_K F = 0$, we must also have that:
\beq
{\mathcal{L}}_K * F = 0 \, .
\eeq
The equations of motion 
of $F$ state that: \beq d * F + \frac{1}{2} F \wedge F = 0. \eeq
So we can introduce a 6-form gauge potential $C$ for $*F$: \beq
\label{dC}dC = *F + \frac{1}{2} A \wedge F \, , \eeq in such a way
that \beq {\mathcal{L}}_K C = 0 \, . \eeq Now consider the 5-form
$\Sigma + \iota_K C + A \wedge (\omega + \frac{1}{2} \iota_K A)$.
This is closed: \beqa d(\Sigma + \iota_K C + A \wedge (\omega +
\frac{1}{2} \iota_K A))
 = \iota_K *F - \omega \wedge F + {\mathcal{L}}_K C - \iota_K(*F + \frac{1}{2} A \wedge F) \nonumber \\
 + \,  F \, \wedge \,(\omega + \frac{1}{2} \iota_K A) - A \wedge (\iota_K F + \frac{1}{2} ({\mathcal{L}}_K A - \iota_K F)) \, = \, 0 \, .
\eeqa
This closed 5-form will be the topological charge for an M5-brane.


\section{Supersymmetry algebras}

\subsection{Spacetime supersymmetry algebras}
\label{spacetime} In this section we review the construction
\cite{Figueroa, Papadopoulos} of the supersymmetry algebra
associated with a solution of eleven dimensional supergravity. As
a starting point, consider the super-isometry algebra of 11d
Minkowski space, i.e.\ the super-Poincar\'{e} algebra. Of particular
relevance to the following discussion will be the subalgebra of
supertranslations generated by the 32 component Majorana spinor
charges $Q_\alpha$. We have:

\beq \label{supertranslations}\{Q_\alpha, Q_\beta \} = (C
\Gamma^M)_{\alpha\beta} P_{M} \eeq and \beq [P_M, Q_\alpha] = 0 \,
, \qquad [P_M , P_N] = 0 \, . \eeq

We can rewrite equation \reef{supertranslations} in an equivalent
way by introducing a commuting Majorana spinor parameter
$\epsilon^\alpha$ and demanding that for arbitrary
$\epsilon^\alpha$:

\beq \{ \epsilon^\alpha Q_\alpha, \epsilon^\beta Q_\beta\} =
(\epsilon^T C\Gamma^M \epsilon) P_M . \eeq
For Majorana spinors, $\epsilon^T C = \bar{\epsilon}$ and so
introducing the vector \beq \label{K} K^M \equiv \bar{\epsilon}
\Gamma^M \epsilon \eeq this becomes: \beq \label{qsquared}
2(\epsilon Q)^2 = K^M P_M. \eeq

Now consider the symmetry algebra of a general solution of eleven
dimensional supergravity. Preserved supersymmetries are given by
commuting Majorana spinor fields $\epsilon^\alpha(x)$ satisfying
the Killing spinor equations \reef{Killing2}: \beq \tilde{D}_M
\epsilon = 0. \eeq

For any such Killing spinor $\epsilon^\alpha(x)$ there is a
corresponding supercharge $ \epsilon Q$. Thus the number of
linearly independent supercharges is given by the dimension of the
space of solutions to the Killing spinor equations. The algebra of
these supercharges is given by equation \reef{qsquared}. Note that
$K$ defined in equation \reef{K} is now  a field also.

We expect $K^M P_M$ to correspond to a bosonic symmetry of the
solution. These are given by infinitesimal coordinate
transformations which leave the solution invariant. An
infinitesimal diffeomorphism is associated with a vector field
acting by the Lie derivative. So bosonic symmetries are associated
with vector fields $K^M(x)$ which obey Killing's equation \beq
{\mathcal L }_K g = 0\eeq and also \beq {\mathcal L }_K F = 0,
\eeq where $F$ is the four-form field strength of 11d
supergravity. As we saw in section 2, it is an automatic
consequence of the Killing spinor equations that $K$, constructed
as in equation \reef{K}, obeys these equations. Note that in
general, there may be other isometries which are not generated by
equation \reef{qsquared}.

Since the Killing vectors $K^M$ generate infinitesimal coordinate
transformations they act on each other by the Lie derivative:
\beq
[K^M P_M , J^N P_N] = ({\mathcal L}_{K} J)^R P_R
\eeq

Under an infinitesimal coordinate transformation which leaves the
metric invariant, the vielbein undergoes a Lorentz rotation and so
spinors and other objects which transform under change of vielbein
are also transformed. Thus the Killing vectors act on spinor
fields by the spinorial Lie derivative\footnote{See e.g.
 \cite{Figueroa} for a definition. }: \beq [K^M P_M,
\epsilon^\alpha Q_\alpha] = ({\mathcal L}_K \epsilon)^\beta
Q_\beta. \eeq

It is a non-trivial consistency check that the super-Jacobi
identities are satisfied as an automatic result of this
construction \cite{Figueroa}.


\subsection{Supersymmetry algebra for membranes}
\label{membranes}
We now consider the addition of branes. Let's
start once again with the super-Poincar\'{e} algebra of flat
eleven dimensional space. As stated in the introduction, if we
couple a supermembrane probe to this background, the resulting
membrane action inherits the symmetries of the background, but
with a modification to the supertranslation algebra
\cite{deAzcarraga}:

\beq \label{brane} \{Q_\alpha, Q_\beta \} = (C
\Gamma^M)_{\alpha\beta} P_{M} \pm \frac{1}{2} (C \Gamma_{M
N})_{\alpha\beta} Z^{M N}, \eeq where \beq Z^{M N} = \int dX^M
\wedge dX^N \eeq and the integration is taken over the spatial
worldvolume of the membrane. Explicitly, if we introduce coordinates $(\sigma_1,
\sigma_2)$ on the spatial worldvolume we have: \beq Z^{M N} =
\frac{1}{2} \int \epsilon^{i j} \frac{\partial X^M}{\partial
\sigma^i} \frac{\partial X^N}{\partial \sigma^j} d\sigma^1 \wedge
d\sigma^2. \eeq Similarly, the momentum $P_M$ involves an
integration over the spatial worldvolume of a momentum density
$p_M(\sigma)$: \beq P_M = \int d^2 \sigma p_M(\sigma). \eeq

As before, we introduce the constant Majorana spinor parameter
$\epsilon^\alpha$ and rewrite equation \reef{brane} as: \beq
2(\epsilon Q)^2 = K^M P_M \pm \omega_{M N} Z^{M N}, \eeq where we
have also introduced the two-form $\omega_{M N}$ defined by: \beq
\label{Omega} \omega_{M N} = \bar{\epsilon} \Gamma_{M N} \epsilon.
\eeq We can write this a little more suggestively by taking the
(constant) parameters $K^M$ and $\omega_{M N}$ inside the
integral: \beq \label{brane2}
 2(\epsilon Q)^2 =
\int d^2 \sigma K^M p_M \pm \int \omega \, . \eeq

A proposal for the generalisation of this formula to membranes in
curved 11d supergravity backgrounds for which $K^M$ is timelike was presented in
\cite{Gutowski}. This generalisation was motivated by
considerations of kappa-symmetry. In our notation the
generalisation of \reef{brane2} to a general curved background (without imposing any restriction on $K^M$) is:
\beq\label{curved}
 2(\epsilon Q)^2 = \int d^2 \sigma K^M p_M \pm \int
 (\omega + \iota_K A).
\eeq In this expression, $A$ is a three-form potential for the
four-form field strength $F$.\footnote{The combination of $K^M P_M
\pm \iota_K A$ is very natural since the brane is electrically
charged under $A$ and this generalises the replacement of $p_\mu$
with $p_\mu + e A_\mu$ for a charged particle in an
electromagnetic field.} Note that $K$ and $\omega$ are no longer
constant, but are the fields built from the Killing spinors
according to equations \reef{K} and \reef{Omega}. Also note, that
a particular gauge choice for $A$ has been taken so that \beq
{\mathcal L}_K A = 0 \, . \eeq This is possible to do since
${\mathcal L}_K F = 0$ and just corresponds to a choice of gauge
potential $A$ which preserves the symmetry generated by $K$. In
order for our proposal to make sense, we require that the second
term in \reef{curved} be topological, i.e.\ \beq d (\omega + \iota_K
A) = 0 \, . \eeq As we saw in section 2, this equation is a
consequence of the Killing spinor equations and the definition of
$\omega$.

The supersymmetry algebra \reef{curved} leads to a BPS type bound
on the energy/momentum of the M2-brane, since $(\epsilon Q)^2 \geq
0$. We find: \beq \int d^2 \sigma K^M p_M \geq \mp \int
 (\omega + \iota_K A), \eeq
where the term on the RHS is topological. This bound was found for
the case of timelike $K^M$ in \cite{Gutowski}.


\subsection{Supersymmetry algebra for five-branes}
\label{fivebranes}
In flat space the supertranslation algebra for
a five-brane probe is: \beq \label{5brane} \{Q_\alpha, Q_\beta \}
= (C \Gamma^M)_{\alpha\beta} P_{M} \pm \frac{1}{5!} (C \Gamma_{M N
P Q R})_{\alpha\beta} Z^{M N P Q R}, \eeq where \beq Z^{M N P Q R}
= \int dX^M \wedge dX^N \wedge dX^P \wedge dX^Q \wedge dX^R \eeq
and the integration is taken over the spatial worldvolume of the
five-brane.

Written in terms of the parameter $\epsilon$ this becomes:
\beq
2(\epsilon Q)^2 = K^M P_M \pm \Sigma_{M N P Q R} Z^{M N P Q R},
\eeq
where we
have also introduced the five-form $\Sigma_{M N P Q R}$ defined by: \beq
\label{Sigma} \Sigma_{M N P Q R} = \bar{\epsilon} \Gamma_{M N P Q R} \epsilon. \eeq

To extend this to a curved spacetime background we let
$\epsilon(x)$ be a Killing spinor field and following the membrane
case we might guess that the algebra becomes: \beq 2(\epsilon Q)^2
= \int d^5 \sigma K^M p_M(\sigma) \pm \int (\iota_K C + \Sigma),
\eeq where C is the gauge potential for $F$. Actually, as we saw
in section 2, this is not closed and so the correct answer is:
\beq 2(\epsilon Q)^2 = \int d^5 \sigma K^M p_M(\sigma) \pm \int
(\iota_K C + \Sigma + A \wedge (\omega + \frac{1}{2} \iota_K A)).
\eeq

It is also interesting to consider the supersymmetry algebra with
non-zero worldvolume gauge field $B$. We can construct a closed
five-form using the closed two-form $\omega + \iota_K A$ and the
closed three-form $dB$. With such a term, the supersymmetry
algebra becomes: \beq \label{nonzerodB}2(\epsilon Q)^2 = \int d^5
\sigma K^M p_M(\sigma) \pm \int (\iota_K C + \Sigma + (A + dB)
\wedge (\omega +  \iota_K A) -  \frac{1}{2} A \wedge \iota_K A).
\eeq This equation agrees with and generalises results in
\cite{Sorokin, Barwald}, for the M5-brane supersymmetry algebra with
non-zero worldvolume fields. \footnote{Note that the coefficient
of the $dB$ term in the previous equation is not fixed by the
requirement that the five-form be closed, since we are adding
together two five-forms which are individually closed. We can fix
the normalisation by reference to the flat space analysis of
\cite{Barwald}.}

Once, again the supersymmetry algebra, \reef{nonzerodB} leads to a
BPS bound on the energy/momentum: \beq \label{BPS} \int d^5 \sigma
K^M p_M(\sigma) \geq \mp \int (\iota_K C + \Sigma + (A + dB)
\wedge (\omega +  \iota_K A) -  \frac{1}{2} A \wedge \iota_K A)
.\eeq


\section{An example: M5-brane background}
\label{examples}

We now present an example to illustrate the general approach for
constructing the superalgebras of brane probes and to point out a
subtlety in our expressions for the topological charges. First, we
need to choose a supersymmetric background in which to work. We
choose to consider the supergravity background corresponding to a
collection of coincident M5-branes. We will also need to choose a
specific supersymmetry of the background and we will choose one
such that $K^2 = 0$.

The metric and 7-form are given by
\begin{eqnarray}
ds^2 &=& H^{-1/3}\left( - (dx^0)^2 + (dx^1)^2 + (dx^2)^2 + (dx^3)^2 + (dx^4)^2 +(dx^5)^2 \right)\nonumber\\
&&+ H^{2/3} \left( (dx^6)^2 + (dx^7)^2 +(dx^8)^2 +(dx^9)^2 +(dx^{\natural})^2 \right)\\
\ast F &=& - d H^{-1} \wedge dx^0 \wedge dx^1\wedge dx^2\wedge
dx^3\wedge dx^4 \wedge dx^5
\end{eqnarray}
where $H$ is a harmonic function which depends on the radial
distance, $r$, from the brane where \[r^2 = (x^6)^2 + (x^7)^2
+(x^8)^2 +(x^9)^2 +(x^{\natural})^2 \, .\]
We also have: \beq F = \frac{1}{r}\ \frac{\partial H }{\partial
r}\ \frac{1}{4!}\ \epsilon_{ijklm} x^{i} dx^{j} \wedge
dx^{k}\wedge dx^{l}\wedge dx^{m} \label{F4} \eeq where $i, j,
\ldots$ run over the indices $\{ 6, 7, 8, 9, \natural\}$.

The background has 16 Killing spinors $\epsilon = H^{-1/12}
\epsilon_0$ which are constructed from constant spinors
$\epsilon_0$ satisfying: \beq \Gamma_{012345}\epsilon_0 =
\epsilon_0  \, .\eeq  Note that if we normalise $\epsilon_0$ such
that $\epsilon_0^T \epsilon_0 = 1$ then $\epsilon^T \epsilon =
H^{-1/6}$.  One can make the projections of equation \reef{null}
on the Killing spinors of the background:
\[\Gamma_{01} \epsilon = - \epsilon\]
\beq \Gamma_{2345}\epsilon = \Gamma_{2367}\epsilon = \Gamma_{2468}
\epsilon = \Gamma_{2389}\epsilon = - \epsilon \eeq These
projections are consistent with the projection condition for the
the background M5-brane: $\Gamma_{012345}\epsilon = \epsilon$.

As discussed in section \ref{closedforms}, the above conditions
define a null Killing vector, $K$. In this background $K$,
$\omega$ and $\Sigma$ are given explicitly by
\begin{eqnarray}
K &=& - H^{-1/3} ( dx^0  + dx^1) \label{Kexplicit}\\
\omega &=& - K\wedge e^{\natural}=  ( dx^0 + dx^1 ) \wedge dx^{\natural}\label{omegaexplicit}\\
\Sigma &=& - K \wedge \phi \label{sigmaexplicit}
\end{eqnarray}
where $\phi$ is the Cayley 4-form given by
\begin{eqnarray}
\phi &=&  H^{-2/3}  dx^2 \wedge dx^3 \wedge dx^4\wedge dx^5 + H^{4/3} dx^6\wedge dx^7 \wedge dx^8 \wedge dx^9 \nonumber\\
&& H^{1/3} \big[ dx^2\wedge dx^3 \wedge dx^6 \wedge dx^7 -  dx^3\wedge dx^4 \wedge dx^7 \wedge dx^8 + dx^2\wedge dx^4 \wedge dx^6 \wedge dx^8 \nonumber \\
&& + dx^3\wedge dx^5 \wedge dx^7 \wedge dx^9 - dx^2\wedge dx^5 \wedge dx^6 \wedge dx^9 + dx^4\wedge dx^5 \wedge dx^8 \wedge dx^9 \nonumber \\
&& + dx^4\wedge dx^5 \wedge dx^6 \wedge dx^7 - dx^3\wedge dx^4 \wedge dx^6 \wedge dx^9 + dx^2\wedge dx^3 \wedge dx^8 \wedge dx^9 \nonumber\\
&& - dx^2\wedge dx^5 \wedge dx^7 \wedge dx^8 - dx^2\wedge dx^4
\wedge dx^7 \wedge dx^9 - dx^3\wedge dx^5 \wedge dx^6 \wedge
dx^8\big]
\end{eqnarray}
We now check that the differential equations for $K, \omega$ and
$\Sigma$, given in Eqs.~(\ref{dK})-(\ref{sigma}) are satisfied.
Given the form of $F$ it is clear that $\iota_K F = \iota_{\omega}
F= 0$.
 Clearly from Eq.~(\ref{omegaexplicit}) we have
\beq d \omega = 0 = \iota_K F \eeq and thus Eq.~(\ref{omega}) for
$\omega$ is satisfied. We now consider the differential equation
(\ref{dK}) for $K$. From the explicit form of $K$ in
Eq.~(\ref{Kexplicit}) we have \beq dK = - \frac{1}{3} H^{-4/3}
(dx^0 + dx^1)\wedge dH \eeq Since $\iota_{\omega} F = 0 $ we
simply have to show that this is equal to $\frac{1}{3}
\iota_{\Sigma}\ast F$. From the form of $\ast F$ it is clear that
the only term in $\Sigma$ that makes a non-zero contribution to
$\iota_{\Sigma}\ast F$ is $-K\wedge H^{-2/3}dx^2 \wedge dx^3
\wedge dx^4\wedge dx^5$, which comes from the first term in the
Cayley 4-form. Therefore we find \beq \iota_{\Sigma} \ast F =
H^{2/3}(dx^0 + dx^1)\wedge dH^{-1} = - H^{-4/3} (dx^0 +
dx^1)\wedge dH = 3 dK \eeq and so Eq (\ref{dK}) is satisfied.
Finally we verify that the differential equation for $\Sigma$,
Eq.~(\ref{sigma}) holds. Now
\begin{eqnarray}
d \Sigma &=&  dH \wedge (dx^0 + dx^1)\wedge \left( - H^{-2}\ dx^2
\wedge dx^3\wedge dx^4\wedge dx^5 +   dx^6 \wedge dx^7 \wedge
dx^8\wedge dx^9\right)\nonumber\\\label{dsigma}
\end{eqnarray}
We can also calculate \beq \iota_K \ast F = - H^{-2} dH  \wedge
(dx^0 + dx^1)\wedge dx^2 \wedge dx^3 \wedge dx^4\wedge dx^5 \eeq
which clearly agrees with the first term in $d\Sigma$. Also
\begin{eqnarray}
\omega \wedge F &=& \frac{\partial H}{\partial r} (dx^0 + dx^1) \wedge \frac{x^{\natural} dx^{\natural}}{r} \wedge dx^6 \wedge dx^7 \wedge dx^8 \wedge dx^9\nonumber\\
&=& \frac{\partial H}{\partial r} (dx^0 + dx^1) \wedge dr \wedge
dx^6 \wedge dx^7 \wedge dx^8 \wedge dx^9\label{owedgef}
\end{eqnarray}
so from Eqs.(\ref{dsigma})-(\ref{owedgef}) we see that \beq d
\Sigma = \iota_K \ast F - \omega\wedge F \eeq as required.

We now find the closed two and five forms appearing in the
membrane and fivebrane superalgebras. We can choose a gauge for
$A$ with $\mathcal{L}_K A = 0$, such that $\iota_K A = 0$ and then
the two form is just $\omega$. The non-zero terms in the five form
are then \beq \iota_K C + \Sigma + (A + dB) \wedge \omega \, .\eeq
We can pick $C$ so that \beq \iota_K C = -(H^{-1} - 1) (dx^0 +
dx^1) \wedge dx^2 \wedge dx^3 \wedge dx^4 \wedge dx^5 \, . \eeq
The remaining subtlety is in how to define $A \wedge \omega$ since
$A$ is a magnetic potential for the background solution and not
globally well-defined. The natural solution in this case is to
define the integral of $A \wedge \omega$ over the 5d spatial
worldvolume of the brane via an integral of $F \wedge \omega$ over
a 6d surface whose boundary is the 5d spatial worldvolume. From
the expression for $F \wedge \omega$ above we see that this can be
simply integrated to give back a 5d integral of \beq -(H - 1)
\wedge (dx^0 + dx^1)\wedge dx^6 \wedge dx^7 \wedge dx^8 \wedge
dx^9 \, . \eeq Putting everything together we find that the
expression for the five form is just \beq  (dx^0 + dx^1) \wedge
\phi_f + dB \wedge \omega \, , \eeq where $\phi_f$ is just the
flat space Cayley four-form. This expression is manifestly closed.
We see that for a choice of supersymmetry which is preserved by
the background brane, the supersymmetry algebra for membrane and
fivebrane probes is unaltered from flat space.

\section{Discussion}
\label{conclusions}

We have presented expressions for the supersymmetry algebras of
membranes and M5-branes in arbitrary supersymmetric backgrounds of
eleven dimensional supergravity. In particular, we have shown how
supersymmetry ensures the existence of closed two and five forms
which appear as topological charges in the algebras. It should be
straightforward to apply the same ideas to other supergravity
theories in different dimensions, giving a simple derivation of
the supersymmetry algebras for the branes in these theories.

One motivation for our work was to understand better the dynamics
of branes which preserve supersymmetries related to null Killing
spinors. Examples of such branes are giant gravitons
\cite{McGreevy,Grisaru,Hashimoto,Mikhailov}, null intersecting
branes \cite{Acharya:planes2,Bachas:null,Myers:from},
supertubes/M-ribbons
\cite{Mateos:supertubes,Emparan:supergravity,Mateos:supercurves,Hyakutake:supertubes}.The
analysis of the supersymmetry algebras and related BPS bounds
which we present here is a useful step towards this understanding,
but there is certainly more to be done here. It would be very
interesting to find a way of describing the most general brane
configuration which preserves the supersymmetry associated with a
particular Killing spinor.

It would also be interesting to extend our analysis to cases in
which there are several Killing spinors and to understand better
the structure of the supersymmetry algebra in those cases. A first
step for doing this would be to classify normal forms for
projection conditions preserving different numbers of
supersymmetries in order to find the algebraic structure of the
associated forms.


\vspace{1cm} \noindent {\bf Acknowledgements}\\
We would like to thank Jerome Gauntlett for helpful comments on an
early draft of the paper. DCP thanks Kazuo Hosomichi, Rob Myers
and Amanda Peet for  discussions and is grateful to the Perimeter
Institute for hospitality during the completion of this paper. DCP
is supported by Ontario PREA. EJH is funded by the University of
Adelaide and the Overseas Research Students Awards Scheme.


\bibliographystyle{utphys}
\bibliography{refs}

\end{document}